\documentclass[aps,prl,twocolumn,superscriptaddress,showpacs]{revtex4-1}

\usepackage{amsmath,amssymb}
\usepackage[utf8x]{inputenc}
\usepackage[english]{babel}
\usepackage{graphicx,epsfig}
\usepackage{xspace}
\usepackage[colorlinks,linkcolor=blue,citecolor=blue]{hyperref}
\usepackage{times}
\usepackage{stmaryrd}
\usepackage[usenames,dvipsnames]{color}
\usepackage[normalem]{ulem}

\usepackage{comment}

\definecolor{dgreen}{rgb}{0.0, 0.5, 0.0}

\definecolor{dgreen}{rgb}{0.0, 0.5, 0.0}

\definecolor{dgreen}{rgb}{0.0, 0.5, 0.0}

\newcommand{\eqn}[1]{
\begin{eqnarray}
	#1
\end{eqnarray}
}

\begin{document}


\title{Many-body interferometry of magnetic polaron dynamics}

\author{Yuto Ashida}
\affiliation{Department of Physics, University of Tokyo, 7-3-1 Hongo, Bunkyo-ku, Tokyo
113-0033, Japan}
\author{Richard Schmidt}
\affiliation{ITAMP, Harvard-Smithonian Center for Astrophysics, Cambridge, Massachusetts 02138, USA}
\affiliation{Department of Physics, Harvard University, Cambridge, Massachusetts 02138, USA}
\author{Leticia Tarruell}
\affiliation{ICFO-Institut de Ci{\`e}ncies Fot{\`o}niques, The Barcelona Institute of Science and Technology, 08860 Castelldefels (Barcelona), Spain}
\author{Eugene Demler}
\affiliation{Department of Physics, Harvard University, Cambridge, Massachusetts 02138, USA}

\date{\today}

\begin{abstract} 
The physics of quantum impurities coupled to a many-body environment is among the most important paradigms of condensed matter physics. In particular, the formation of polarons,  quasiparticles dressed by the polarization cloud, is key to the understanding of transport, optical response, and induced interactions in a variety of materials.
Despite recent remarkable developments in ultracold atoms and solid-state materials, the direct measurement of their ultimate building block, the polaron cloud, has remained a fundamental challenge. 
We propose and anlalyze a unique platform to probe time-resolved dynamics of polaron-cloud formation with an interferometric protocol. We consider an impurity atom immersed in a two-component Bose-Einstein condensate, where the impurity generates spin-wave excitations that can be directly measured by the Ramsey interference of surrounding atoms. The dressing by spin waves leads to the formation of magnetic polarons and reveals a unique interplay between few- and many-body physics that is signified by single- and multi-frequency oscillatory dynamics corresponding to the formation of many-body bound states. Finally, we discuss concrete experimental implementations in ultracold atoms.
\end{abstract}


\maketitle

Understanding the role of interactions between an impurity and its environment  is a fundamental problem in quantum many-body physics. A central concept for the description of such systems is a "dress" of collective excitations surrounding the impurity, also  known as the polaron cloud \cite{Mahan00}. It crucially determines thermodynamic and transport properties of a wide variety of condensed matter systems including doped semiconductors \cite{Jungwirth06}, metallic ferromagnets \cite{Majumdar98}, high-temperature superconductors \cite{Salje95}, $^3$He-$^4$He mixtures \cite{Bardeen1967}, and perovskites \cite{Teresa97}.
Meanwhile, recent experimental realizations of imbalanced mixtures of ultracold atoms have opened up new possibilities for studying polaron physics in a highly controlled manner.  Until now most studies focused on impurities interacting with a single-component Bose-Einstein condensate (BEC) \cite{Mathey2004,Palzer2009,WC12,Cucchietti2006,Klein2007,Tempere2009,Casteels2011pra,Spethmann2012,Casteels2013,Rath2013,Li2014,Christensen2015,Giorgini2015,Giorgini2016,Levinsen2015,HMG16,JNB16,Shchadilova2016,RG,Vlietinck2015,SchmidtLem2015,Volosniev2015,Shchadilova2016pra,SchmidtLem2016,SchmidtDem2016,bellotti2016,Midya2016} or 
Fermi gas of atoms \cite{Schirotzek2009,NS09,CX10,schmidt_excitation_2011,massignan_repulsive_2011,Schmidt2012b,kohl,grimm,Zhang2012,Mathy2012,PM14,CX15,Ong2015,Meinert2016}. This allowed  to take first steps to explore physics of polaron beyond the Fr\"ohlich paradigm and the Anderson orthogonality catastrophe \cite{Knap13,Cetina2015,Cetina2016,Schmidt2017,SF16}. Despite these remarkable developments, measuring its ultimate building block, the polaron cloud, has remained a  challenge not only in ultracold atoms but also in solid-state materials. A major difficulty stemmed from the elusive nature of polaron cloud as it is associated with subtle density change in  the environment arising from the interaction with the impurity.

\begin{figure}[b!]
\includegraphics[width=86mm]{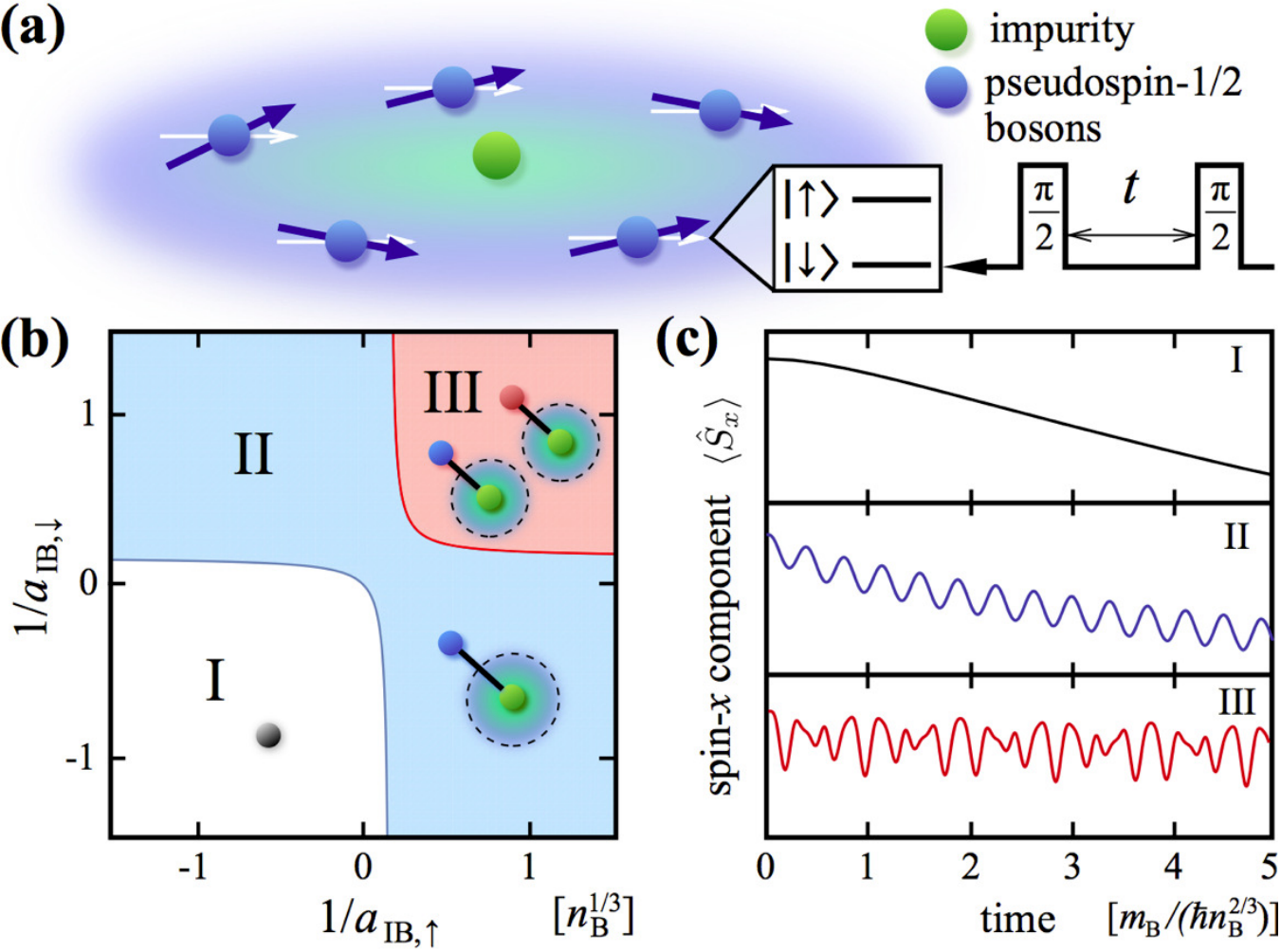} 
\caption{\label{fig1}
(color online).  (a) Illustration of an impurity atom immersed in a two-component BEC. Arrows indicate the internal pseudospin states of BEC atoms. By applying a $\pi/2$-pulse, the host bosons are initially prepared in an equal superposition of  spin $\uparrow$- and $\downarrow$-state. The interaction between the impurity and host atoms induces spin dephasing, which is measurable by Ramsey interferometry.
(b) Depending on the impurity-boson scattering  lengths $a_{{\rm IB},\uparrow\downarrow}$, the system is characterized by the existence of zero, one, and two many-body bound states (regions I, II, and III). The diagram is plotted for a boson-boson scattering length $a_{\rm BB}n_{\rm B}^{1/3}=0.05$, and $m_{\rm I}/m_{\rm B}=0.95$ as appropriate for a mixture of $^{41}$K-$^{39}$K atoms. (c) The BEC spin dephasing dynamics exhibits (I) monotone relaxation, (II) single- and (III) multi-frequency oscillations reflecting the existence of bound states. The dephasing signal is proportional to the number of impurities, which fixes the scale of the vertical axis.
}
\end{figure}

In this Rapid Communication, we show that the use of Ramsey interferometry performed on \textit{bath atoms} can overcome the challenge and allows a direct measurement of polaron-cloud formation in real time. Applying it to impurity atoms immersed in a two-species Bose-Einstein condensate (BEC), we analyze impurities interacting with a magnetic environment and study the impact of polaron-cloud formation on the many-body environment. The setup is illustrated in Fig.~\ref{fig1}(a); the host BEC atoms provide an artificial ferromagnetic medium in which the impurity is dressed by spin-wave excitations, leading to the formation of a magnetic polaron. In previous setups \cite{Mathey2004,Palzer2009,WC12,Cucchietti2006,Klein2007,Tempere2009,Casteels2011pra,Spethmann2012,Casteels2013,RG,Vlietinck2015,SchmidtLem2015,Christensen2015,Volosniev2015,Shchadilova2016pra,SchmidtLem2016,Rath2013,Li2014,Giorgini2015,Levinsen2015,SchmidtDem2016,Shchadilova2016}, the impurity is coupled only to phonon excitations and observing its cloud formation poses a daunting challenge due to the difficulty to measure a minuscule density change around the impurity. In contrast, a magnetic polaron is dressed by a spin-polarized cloud created from changes in spin configurations \cite{deGennes60}. It is this magnetic dressing that enables one to directly measure the polaron cloud by performing Ramsey interferometry on \textit{surrounding atoms}, revealing its rich out-of-equilibrium dynamics. As a striking feature that is not readily attainable in solid-state systems, we find that the polaron cloud is composed of many-body bound states in the strong-coupling regime. This leads to a unique `phase diagram' of the polaron cloud (Fig.~\ref{fig1}(b)), which characterizes distinct oscillatory real-time dynamics in the many-body environment (Fig.~\ref{fig1}(c)). 
Moreover, our scheme can effectively enhance signal amplitudes from the impurity because the impurity creates multiple excitations in the bath that can be directly detected in experiments. This novel protocol can be transferred to a multitude of experimental systems \cite{MA16,VG17,AVG09,MGE14,ZJ16,FB16,DP17} in which interferometric schemes are readily available. Our approach thus implies possibilities for enhancing the detectability of impurity physics in a way different from previous studies, where the impurity itself was probed either by radio-frequency  \cite{Schirotzek2009,NS09,kohl,grimm,Zhang2012,WC12,HMG16,JNB16} or interferometric measurements \cite{Cetina2015,Cetina2016,Parish2016,Schmidt2017} and thus the signal amplitudes were intrinsically limited by the number of impurities.

{\it Model.---}
We consider an impurity of mass $m_{\rm I}$ having no internal degrees of freedom and being immersed in a weakly interacting \textit{two-component}  spinor BEC of atoms of mass $m_{\rm B}$ (Fig.~\ref{fig1}(a)).   The system is described by the Hamiltonian

\eqn{\label{H}
\hat{H}=\hat{H}_{\rm B}+\hat{V}_{\rm IB}+\hat{H}_{\rm I},
}
where 
\eqn{\hat{H}_{\rm B}\!=\!\sum_{{\bf k}\sigma}\!\epsilon_{\bf k}\hat{a}_{{\bf k}\sigma}^{\dagger}\!\hat{a}_{{\bf k}\sigma}\!\!+\!\frac{g_{\rm BB}}{2V}\!\!\!\sum_{{\bf k}{\bf k}'{\bf q}\sigma\sigma'}\!\!\!\hat{a}_{{\bf k}\!+\!{\bf q}\sigma}^{\dagger}\!\hat{a}_{{\bf k}'-{\bf q}\sigma'}^{\dagger}\hat{a}_{{\bf k}'\sigma'}\hat{a}_{{\bf k}\sigma}
}
accounts for the background BEC of density $n_{\rm B}$.  The interaction between the impurity and the host bosons is given by
\eqn{\hat{V}_{\rm IB}=\frac{1}{V}\sum_{{\bf k}{\bf q}\sigma}g_{{\rm IB},\sigma}\hat{a}_{{\bf k}+{\bf q}\sigma}^{\dagger}\hat{a}_{{\bf k}\sigma}e^{i{\bf q}\hat{{\bf R}}},
}
and $\hat{H_{\rm I}}=\hat{\bf P}^2/(2m_{\rm I})$ is the kinetic energy of the impurity.
The operators $\hat{a}_{{\bf k}\sigma}$ ($\hat{a}^{\dagger}_{{\bf k}\sigma}$) annihilate (create) the host bosons with wavenumber $\bf k$ and spin $\sigma=\uparrow,\downarrow$, and $\epsilon_{\bf k}=\hbar^2{\bf k}^2/(2m_{\rm B})$ is their dispersion relation. The momentum (position) of the impurity is described by $\hat{\bf P}$ ($\hat{\bf R}$). We assume spin independent interactions between the host bosons characterized by the single parameter $g_{\rm BB}$ as realized for many bosonic species \cite{Kempen02,SC00}. In contrast, the interaction between the impurity and the bosons is spin dependent  and given by $g_{{\rm IB},\sigma}$, which are related to the scattering lengths $a_{{\rm IB},\sigma}$ by the Lippmann-Schwinger equation \cite{SM1}. 

To realize an effective magnetic environment, we initially prepare a superposition state of the pseudospin-1/2 BEC: $|\Psi_{\rm BEC}\rangle\propto(\hat{a}^{\dagger}_{{\bf 0}\uparrow}+\hat{a}^{\dagger}_{{\bf 0}\downarrow})^{N_{\rm B}}|0\rangle$, with $N_{\rm B}$ being the number of host bosons. Due to the SU(2) symmetry of $\hat{H}_{\rm B}$, the internal dynamics of the background bosons of homogeneous density causes no decoherence. In contrast, scattering with the impurity breaks this symmetry and induces  spin dephasing of the medium.
Dealing with a two-component BEC, the collective excitations in the bath correspond not only to  phonon (`charge') excitations ---as in the case of a single-component BEC--- but also spin-wave (`magnon') excitations.  The generation of the latter leads to spin dephasing of the medium or, equivalently, dressing of the impurity by magnons. Following the standard procedure of transforming to the frame comoving with the polaron \cite{Lee1953}, we obtain the effective Hamiltonian \cite{SM1}:

\eqn{\label{calH}
\hat{\cal H}&=&g_{\rm IB}^{+}n_{\rm B}+\frac{(\hat{{\bf P}}-\hat{{\bf P}}_{\rm B})^{2}}{2m_{\rm I}}+\sum_{{\bf k}}\left(\epsilon_{{\bf k}}^{c}\hat{\gamma}_{{\bf k}}^{\dagger c}\hat{\gamma}_{{\bf k}}^{c}+\epsilon_{{\bf k}}^{s}\hat{\gamma}_{{\bf k}}^{\dagger s}\hat{\gamma}_{{\bf k}}^{s}\right)\nonumber\\
&\!+&\!\sqrt{\frac{n_{\rm B}}{V}}\!\sum_{{\bf k}}\!\left[g_{\rm IB}^{+}W_{{\bf k}}\!\left(\!\hat{\gamma}_{{\bf k}}^{c}\!+\!\hat{\gamma}_{-{\bf k}}^{\dagger c}\!\right)\!+\!g_{\rm IB}^{-}\!\left(\hat{\gamma}_{{\bf k}}^{s}\!+\!\hat{\gamma}_{-{\bf k}}^{\dagger s}\!\right)\right]\nonumber\\
&+&\frac{g_{\rm IB}^{+}}{2V}\sum_{{\bf k},{\bf k}'}\left(V_{{\bf k}{\bf k}'}^{(1)}\hat{\gamma}_{{\bf k}}^{\dagger c}\hat{\gamma}_{{\bf k}'}^{c}+\hat{\gamma}_{{\bf k}}^{\dagger s}\hat{\gamma}_{{\bf k}'}^{s}+V_{{\bf k}{\bf k}'}^{(2)}\hat{\gamma}_{{\bf k}}^{\dagger c}\hat{\gamma}_{{\bf k}'}^{\dagger c}+{\rm H.c.}\right)\nonumber\\
&+&\frac{g_{\rm IB}^{-}}{V}\sum_{{\bf k},{\bf k}'}\left(u_{{\bf k}}\hat{\gamma}_{{\bf k}'}^{\dagger s}\hat{\gamma}_{{\bf k}}^{c}-v_{{\bf k}}\hat{\gamma}_{{\bf k}'}^{\dagger s}\hat{\gamma}_{{\bf k}}^{\dagger c}+{\rm H.c.}\right).
}
Here, $\hat{\bf P}$ in this frame represents the total momentum of the system (we consider the case $\hat{\bf P}=0$ hereafter), $\hat{{\bf P}}_{B}=\sum_{{\bf k}}\hbar{\bf k}(\hat{\gamma}_{{\bf k}}^{\dagger c}\hat{\gamma}_{{\bf k}}^{c}+\hat{\gamma}_{{\bf k}}^{\dagger s}\hat{\gamma}_{{\bf k}}^{s})$, $\epsilon_{\bf k}^{c}=\sqrt{\epsilon_{\bf k}(\epsilon_{\bf k}+2g_{\rm BB}n_{\rm B})}$ and $\epsilon_{\bf k}^{s}=\hbar^{2}{\bf k}^2/(2m_{\rm B})$ are the total boson momentum and the dispersion relations of the charge and  spin wave excitations, respectively. The excitations are annihilated (created) by the operators, $\hat{\gamma}_{\bf k}^{c,s}$ ($\hat{\gamma}^{\dagger c,s}_{\bf k}$) that obey the commutation relations $[\hat{\gamma}^{\xi}_{\bf k},\hat{\gamma}^{\eta}_{{\bf k}'}]=[\hat{\gamma}^{\dagger \xi}_{\bf k},\hat{\gamma}^{\dagger \eta}_{{\bf k}'}]=0$, and $[\hat{\gamma}^{\xi}_{\bf k},\hat{\gamma}^{\dagger\eta}_{{\bf k}'}]=\delta_{\xi,\eta}\delta_{{\bf k},{\bf k}'}$ with $\xi,\eta=s,c$. We introduce the vertices $W_{\bf k}=\sqrt{\epsilon_{\bf k}/\epsilon_{\bf k}^{c}}$,  $V_{{\bf k}{\bf k}'}^{(1)}\pm V_{{\bf k}{\bf k}'}^{(2)}=(W_{\bf k}W_{{\bf k}'})^{\pm1}$, as well as $u_{\bf k}$ and $v_{\bf k}$ as the coefficients of the Bogoliubov transformation.  We also introduce the  average and difference of the interaction parameters as $g_{\rm IB}^{\pm}=(g_{{\rm IB},\uparrow}\pm g_{{\rm IB},\downarrow})/2$.  When $g_{\rm IB}^{-}\neq 0$, the imbalance in the impurity-boson interactions switches on spin-charge interactions and generates spin waves.  While the first two lines in Eq.~\eqref{calH} describe the generalization of  Fr{\"o}hlich polaron-type physics \cite{Frohlich1954} to magnon dynamics, the last two lines account for the strong coupling physics that leads to the formation of magnetic polaron bound states.

{\it Interferometry.---} The real-time dynamics of the polaron cloud can be probed through  Ramsey interference of bath atoms. 
Starting with a bath in the $\uparrow$ state, a first $\pi/2$ pulse is used to prepare a superposition of $\uparrow$ and $\downarrow$ as described above.
After the system has evolved for a time $t$, an additional $\pi/2$ pulse is applied and the population $N_{\uparrow}$ of bath atoms remaining in the $\uparrow$ state after the Ramsey protocol is measured. This directly gives the value of the spin-excitation number $N^{s}(t)=\sum_{\bf k}\langle\hat{\gamma}^{\dagger s}_{\bf k}\hat{\gamma}^{s}_{\bf k}\rangle$. One can thus explicitly determine the number $N^{s}(t)$ of spin-wave excitations in the polaron cloud by measuring the atomic population $N_{\uparrow}$  after the Ramsey sequence. 
While we so far analyze the case of a single impurity, our results can be applied to a finite density of impurities as long as the impurity density is sufficiently low such that impurity-impurity interactions remain negligible, as realizable in experimental systems. The interferometric signal $N^{s}=N_{\uparrow}$ is then proportional to the number of impurities and can take a value approaching even a substantial fraction of bath atoms such that it is readily detectable with the current techniques \cite{Cetina2015,Cetina2016,Parish2016,Schmidt2017}. This allows a precise and time-resolved determination of the number of excitations generated in the polaron cloud, which has been challenging to achieve in the previous setups \cite{Mathey2004,Palzer2009,WC12,Cucchietti2006,Klein2007,Tempere2009,Casteels2011pra,Spethmann2012,Casteels2013,RG,Vlietinck2015,SchmidtLem2015,Christensen2015,Volosniev2015,Shchadilova2016pra,SchmidtLem2016,Rath2013,Li2014,Giorgini2015,Levinsen2015,SchmidtDem2016,Shchadilova2016}.



{\it Quantum spin dynamics.---} 
The formation of  magnetic polarons leads to distinct quantum dynamics of the polaron cloud. To demonstrate this, we invoke the time-dependent variational approach \cite{Jackiw1979}.  In particular, we employ a projection onto the submanifold of the Hilbert space spanned by the product of coherent states
\eqn{\label{coh}
|\Psi(t)\rangle=e^{\sum_{{\bf k}}\left(\alpha_{{\bf k}}^{c}(t)\hat{\gamma}_{{\bf k}}^{c}+\alpha_{{\bf k}}^{s}(t)\hat{\gamma}_{{\bf k}}^{s}-{\rm h.c.}\right)}|0\rangle,
}  
where $\alpha_{\bf k}^{c,s}(t)$ are the time-dependent amplitudes of the charge and spin excitations and $|0\rangle$ is their vacuum. The state (\ref{coh}) gives the exact solution for an impurity of infinite mass  immersed into an ideal BEC regardless of the interaction strength between the impurity and  host bosons. 
The equations of motion for $\alpha_{\bf k}^{c,s}$ are given by the variational condition $\delta[\langle\Psi|i\hbar\partial_{t}-\hat{\cal H}|\Psi\rangle]=0$, which results in the coupled integral equations \cite{SM1}:
\eqn{\label{eqmotion}
i\hbar\partial_{t}
\begin{pmatrix}
\alpha^{c}\\
\alpha^{s}
\end{pmatrix}
={\cal M}
\begin{pmatrix}
\alpha^{c}\\
\alpha^{s}
\end{pmatrix}
+{\cal F},
}
where the matrix $\cal M$ and vector $\cal F$ are independent of $\alpha_{\bf k}^{c,s}$. The stationary solution $|\Psi_{\rm mpol}\rangle$ of Eq.~\eqref{eqmotion} contains non-zero spin and charge excitations and represents the magnetic dressed polaron with energy $E_{\rm mpol}=\langle\Psi_{\rm mpol}|\hat{\cal H}|\Psi_{\rm mpol}\rangle$. 

\begin{figure}[b]
\includegraphics[width=83mm]{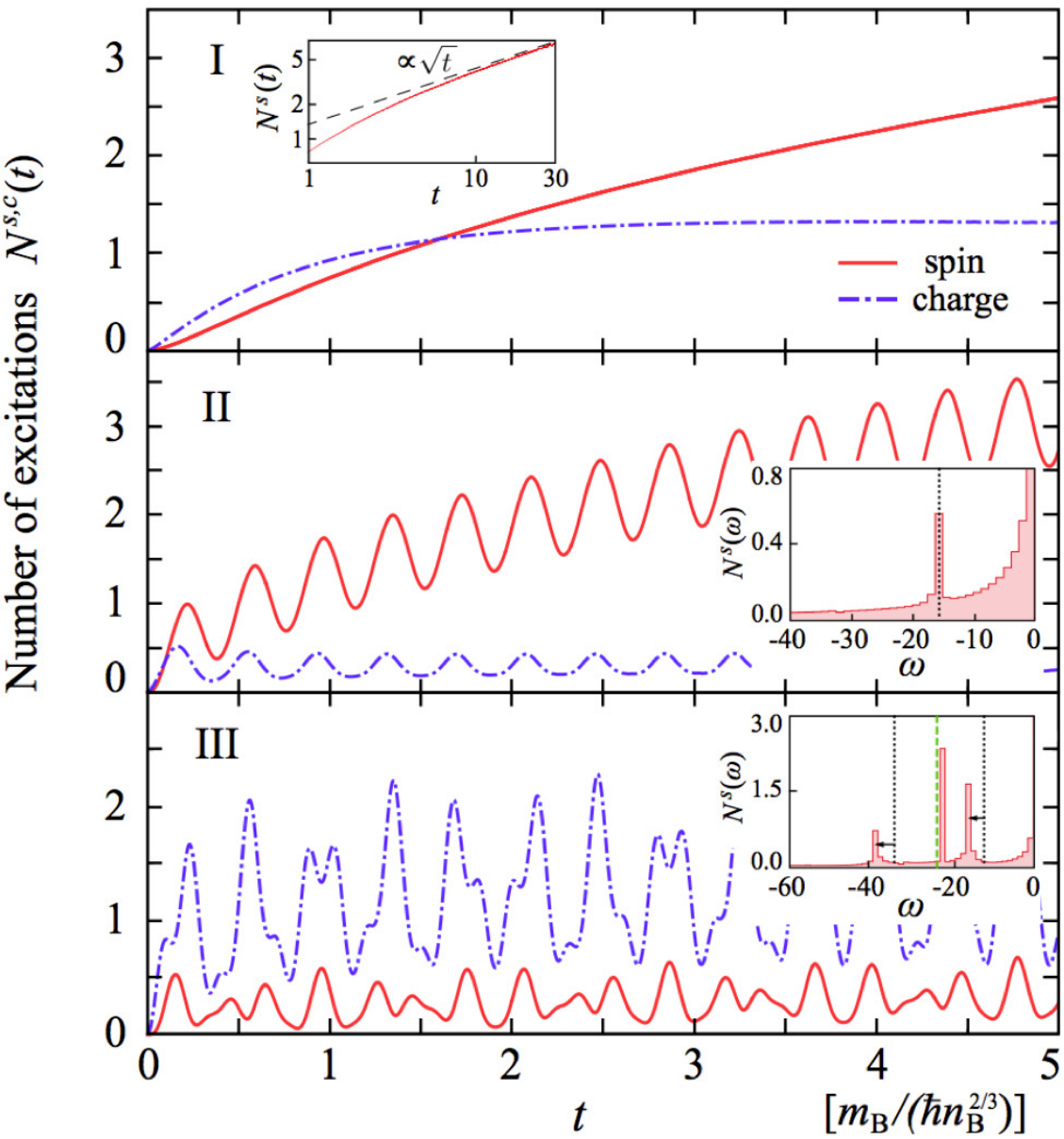} 
\caption{\label{fig2}
(color online).  Quantum dynamics of bath spin (red solid line) and charge (blue dashed line) excitations for a single impurity immersed in a two-component BEC. The impurity-boson scattering lengths are $n_\text{B}^{1/3}(a_{{\rm IB},\uparrow},a_{{\rm IB},\downarrow})=(-1,-10)\;{\rm in\;I},\;(3.5,-5) \;{\rm in\;II},{\; \rm and\;}(3.5,5)\;{\rm in\;III}.$ The inset in I shows the long-time behavior of spin excitations approaching an asymptotic $\sqrt{t}$ scaling. The insets in II and III show the Fourier spectra of $N^{s}(t)$. The energies of the underlying bound states and their difference, calculated from Eq.~(\ref{eigeq}), are shown as black dotted lines and green dashed lines, respectively.
}
\end{figure}

Figure~\ref{fig2} shows the number of spin and charge excitations $N^{s,c}(t)$ for different scattering lengths $a_{{\rm IB},\sigma}$ in the regions I, II, and III, where the system supports zero, one, and two bound states (Fig.~\ref{fig1}(b)).  In the absence of bound states, while $N^{c}(t)$ eventually saturates, $N^{s}(t)$ grows as $\propto\sqrt{t}$ and easily exceeds one (panel I in Fig.~\ref{fig2}).  As a consequence, the observable signals $N^{s}=N_{\uparrow}$ can significantly surpass the number of impurities. In contrast, in the conventional measurements acting on the impurity \cite{Schirotzek2009,NS09,kohl,grimm,Zhang2012,WC12,HMG16,JNB16,Cetina2015,Cetina2016,Parish2016}, the number of detectable signals are strictly limited by that of impurities.
In this regard, interferometric probes acting on the environment can provide a new way to effectively enhance experimental signatures of impurities.

The unbounded generation of spin waves originates from the quadratic nature of the magnon dispersion relation \cite{SM1}. Importantly, collective excitations having quadratic low-energy dispersion ubiquitously appear in many other setups such as fermionic gases \cite{MA16,VG17}, multi-component Bose-Einstein condensates \cite{AVG09,MGE14}, and Rydberg or dipolar gases \cite{ZJ16,FB16,DP17}. This implies a wide applicability of our protocol because interferometric tools of atomic spectroscopy are readily available in these vastly different systems. 

{\it Magnetic dressed bound states.---}
The presence of bound states triggers single- and multi-frequency oscillations in the number of the spin and charge excitations (panels II and III in Fig.~\ref{fig2}). These two different oscillatory dynamics reflect the formation and coupling of many-body bound states, respectively.  To gain further insights, we consider the variational wavefunction

\eqn{\label{bound}
|\Psi_{{\rm b}}(t)\rangle=\sum_{{\bf k}}\left(\psi_{{\bf k}}^{c}(t)\hat{\gamma}_{{\bf k}}^{\dagger c}+\psi_{{\bf k}}^{s}(t)\hat{\gamma}_{{\bf k}}^{\dagger s}\right)|\Psi_{{\rm mpol}}\rangle,
}
which accounts for bound states consisting of single spin-charge excitations bound to the magnetic polaron, i.e., the collective object of the impurity dressed by surrounding many-body excitations. We determine the eigenmodes $\psi_{\bf k}^{c,s}\propto e^{-i\omega t}$ of the equation of motion for the state (\ref{bound}) which yields the eigenvalue equation \cite{SM1}

\eqn{\label{eigeq}
(a_{\rm IB}^{-})^{2}=\left[a_{\rm IB}^{+}\!-\!l_{s}(\omega,a_{\rm IB}^{+})\right]\!\left[a_{\rm IB}^{+}-l_{c}(\omega,a_{\rm IB}^{+})\right].
}
Here $a_{\rm IB}^{\pm}=(a_{\rm IB,\uparrow}\pm a_{\rm IB,\downarrow})/2$ and $1/l_{s,c}=(2\pi/m_{\rm red})(2m_{\rm red}\sum_{\bf k}(1/{\bf k}^2)+\hbar^2\Pi_{s,c})$, which, together with
\eqn{
\Pi_{s}\!\!=\!\!\sum_{\bf k}\!\frac{1}{\hbar\omega\!\!-\!\!E_{\rm mpol}\!\!-\!\!\Omega_{\bf k}^{s}},\,\Pi_{c}\!\!=\!\!\sum_{\bf k}\!\frac{(W^2_{\bf k}\!+\!W^{-2}_{\bf k})/2}{\hbar\omega\!\!-\!\!E_{\rm mpol}\!\!-\!\!\Omega_{\bf k}^{c}},
}
are fully regularized expressions ($\Omega_{\bf k}^{s,c}=\hbar^{2}{\bf k}^{2}/(2m_{\rm I})+\epsilon_{\bf k}^{s,c}$). 
Depending on the scattering lengths $a_{{\rm IB},\sigma}$, Eq.~\eqref{eigeq} has zero, one, or two solutions determining the phase boundaries in  Fig.~\ref{fig1}(b). These boundaries are modified with respect to the corresponding two-body problem as a result of the many-body character of the bound states. In the two-particle problem, a dimer bound state of energy  $\epsilon_{\rm dim}=\hbar^2/(2m_\text{red}a_{{\rm IB},\sigma}^2)$ ($m_{\rm red}=m_{\rm I}m_{\rm B}/(m_{\rm I}+m_{\rm B})$) exists for each positive scattering length $a_{{\rm IB},\sigma}$. As a result,  there are four distinct regions corresponding to the presence or absence of each bound state, i.e., the impurity bound to a host $\uparrow$- or $\downarrow$-boson. Remarkably, the many-body phase diagram in Fig.~\ref{fig1}(b) does not show the corresponding four distinct regimes. Instead, the exchange of magnetic excitations hybridizes the bound states with the medium, resulting in a unified region II.

In this region II, we find that the oscillation frequency governing the bath-spin dynamics agrees with the bound-state energy  calculated from Eq.~(\ref{eigeq}) (inset of panel II in Fig.~\ref{fig2}). In contrast, in region III, the bath-induced coupling between the two bound states manifests itself as a shift in the oscillation frequencies from the bare bound-state energies and also as a large peak at the difference of the two energies  (inset of panel III in Fig.~\ref{fig2}). This effect can be understood as a polaronic nonlinearity introduced by the magnetic medium \cite{SA97}, which induces strongly coupled oscillators dynamics, analogous to polariton-polariton interactions \cite{GNA07} and competing orders in strongly correlated electrons \cite{PS09}.  As we depart from the strongly interacting regime, the coupling of the two bound states weakens and the oscillation frequencies eventually converge to the bare bound-state energies given by Eq.~(\ref{eigeq}) \cite{SM1}.

\begin{figure}[t]
\includegraphics[width=80mm]{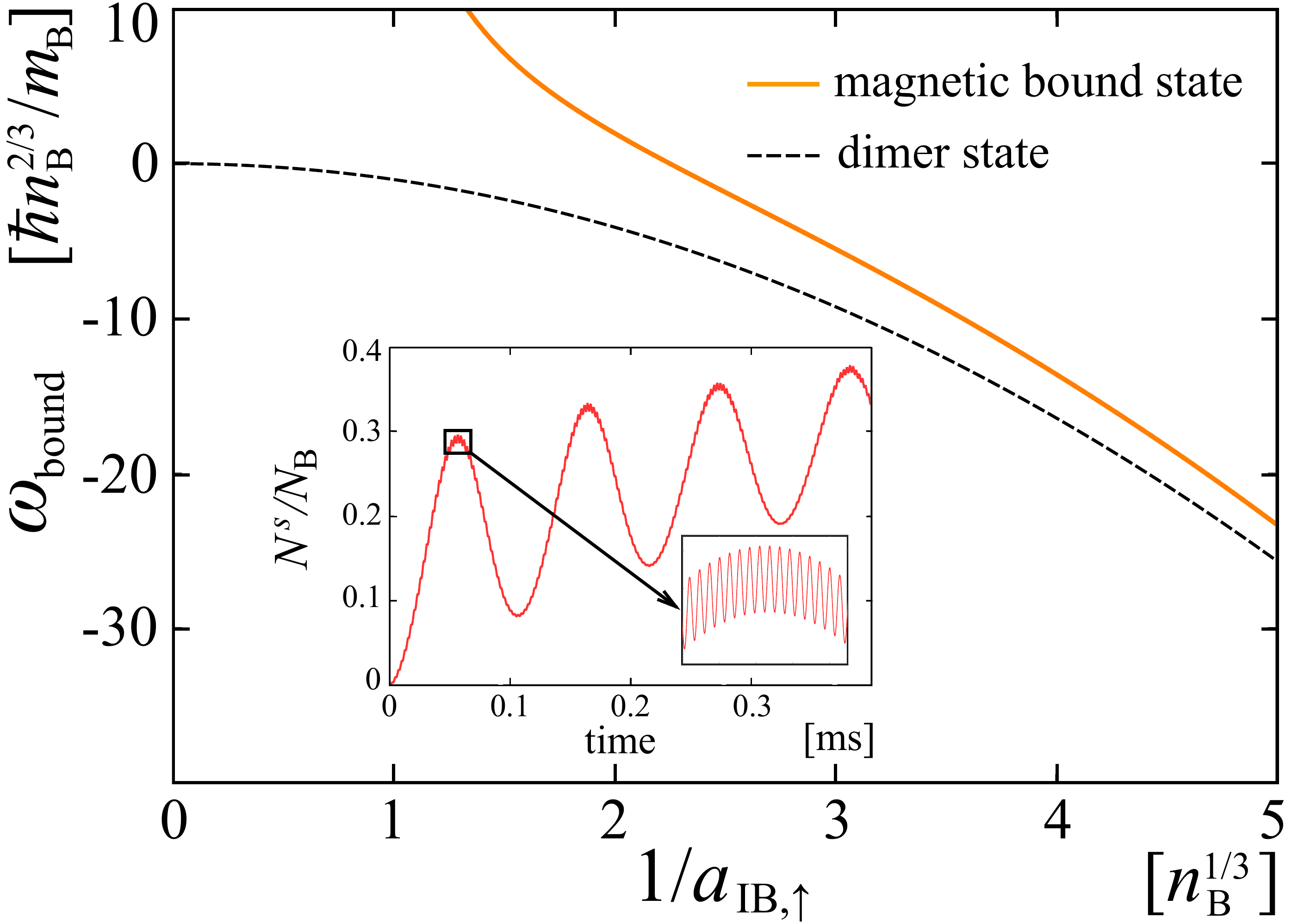} 
\caption{\label{fig3}
(color online).  The magnetic bound-state energy $\omega_{\rm bound}$ (solid line) calculated from Eq.~(\ref{eigeq}), and the bare dimer energy $\epsilon_{\rm dim}=\hbar^2/(2m_\text{red}a_{{\rm IB},\uparrow}^2)$ (dashed line) are plotted against the inverse impurity-bath scattering length $1/a_{{\rm IB},\uparrow}$. The inset shows the dynamics of spin excitations at $1/(a_{{\rm IB},\uparrow}n_{\rm B}^{1/3})=2$ where we assume a homogeneous density $n_{\rm B}=10^{14}\:{\rm cm}^{-3}$, and  a finite impurity density $n_{\rm I}/n_{\rm B}=0.1$. We choose $a_{\rm BB}n_{\rm B}^{1/3}=0.05$, and set $1/(a_{{\rm IB},\downarrow}n_{\rm B}^{1/3})=15$ and $m_{\rm I}/m_{\rm B}=0.95$ as appropriate for  a $^{41}$K-$^{39}$K mixture. 
}
\end{figure}

{\it Experimental implementation.---}
A large number of Bose-Bose and Bose-Fermi mixtures allow for the observation of magnetic polaron physics. As one possible example, we consider here a Bose-Bose mixture of $^{41}$K-$^{39}$K atoms. We identify two miscible states $|\!\!\!\uparrow\rangle=|F=1,m_{F}=1\rangle$ and $|\!\!\!\downarrow\rangle=|F=1,m_{F}=0\rangle$ of $^{41}$K as the host bosons and $|i\rangle=|F=1,m_{F}=1\rangle$ of $^{39}$K as the impurity. In this case, the $|\!\!\uparrow\rangle-|i\rangle$ interaction can be tuned using a Feshbach resonance at 500~G \footnote{Note that our calculation does not take into account finite range effects, which might become important for this particular resonance.}. The imbalance in the scattering lengths of the  two-component host bosons is less than 0.4\% \cite{TomzaPriv}. While such a small breaking of the SU(2) symmetry can in general induce  decoherence of the atomic spins, we confirmed that the effect is negligible compared with the spin dynamics induced by the impurities \cite{SM1}. 

In Fig.~\ref{fig3}, we plot the energy $\omega_{\rm bound}$ of the magnetic-dressed bound state as calculated from Eq.~(\ref{eigeq}). The result is shown in the vicinity of a Feshbach resonance where $a_{{\rm IB},\uparrow}$ takes a large positive value, while $a_{{\rm IB},\downarrow}$ is determined by a  small, positive background value. As shown in the inset of Fig.~\ref{fig3}, this significant imbalance in scattering lengths creates a large number of observable excitations $N^{s}(t)$ in the bath, which can exceed the number of impurities. Here we note that the large spin-excitation number $N^{s}(t)$ is a direct measurable quantity in the proposed Ramsey protocol and can reach to $O(10^4)$ for a typical number $N_{\rm B}\!=\!10^5$ and small, relative impurity density $n_{\rm I}/n_{\rm B}=0.1$. Moreover, the underlying shallow bound state triggers oscillatory dynamics whose frequency is characterized by the bound-state energy that can be typically $\sim\!\!10\:{\rm kHz}$, which should be detectable in the time resolution realized in the current experiments \cite{Cetina2016}. While this oscillation frequency corresponds to a temperature scale $T/k_{\rm B}\!\simeq \!500\:{\rm nK}$ that has been already achieved in several experiments \cite{Cetina2016,FF16,FRJ17}, we note that our predictions should be accessible in higher temperatures by, for example, localizing the impurities around the center of the system \cite{Cetina2016} or by performing local measurements \cite{SF16}. 
We note that there are also other candidates for bath atoms such as $^{87}$Rb and $^{23}$Na, where the imbalance in the scattering lengths can be small enough to observe the predicted phenomena \cite{Kempen02,SC00,SM1}.

{\it Conclusions and Outlook.---}
We showed that the real-time dynamics of the polaron cloud can be directly probed by employing many-body Ramsey interferometry of bath atoms around the impurity. Analyzing an impurity immersed in a two-component Bose gas, we demonstrated that the generation of spin excitations is the key signature of magnetic-polaron formation and found  unique out-of-equilibrium dynamics in the strong-coupling regime such as the characteristic oscillatory behavior governed by the underlying bound states. 
Our protocol acting on the environment rather than the impurity itself can effectively enhance signal amplitudes of impurities owing to a generation of multiple observable excitations (per impurity) in the environment.
This leads to a novel route for observing few-body physics beyond conventional spectroscopy \cite{WC12,HMG16,JNB16} and loss measurements \cite{NP16rev} whose signal amplitudes are intrinsically limited by the number of impurities.
A generalization to large spin spinor BECs \cite{SK13} and the use of $\it in$-$\it situ$ imaging techniques \cite{BWS09,SJF10,YA15,YA16,AA15} can provide new insights in polaron physics. It remains an open question to clarify the role of magnon-mediated interaction  \cite{naidon2016},  potentially leading to an instability of fermionic gases.

\paragraph*{Acknowledgements.---} We acknowledge F. Grusdt, Y. Shchadilova, M. Tomza, M. Ueda, G. Zar{\'a}nd for fruitful discussions.   The authors acknowledge support from the NSF Grant No.~DMR-1308435, Harvard-MIT CUA, AFOSR New Quantum Phases of Matter MURI, the ARO-MURI on Atomtronics, ARO MURI Quism program.  Y.A. acknowledges support from the Japan Society for the Promotion of Science through Program for Leading Graduate Schools (ALPS) and Grant No.~JP16J03613, and Harvard University for hospitality, where this work was completed. R.S. is supported by the NSF through a grant for the Institute for Theoretical Atomic, Molecular, and Optical Physics at Harvard University and the Smithsonian Astrophysical Observatory. L.T. acknowledges support from Fundaci\'{o} Privada Cellex, Spanish MINECO (FIS2014-59546-P and SEV-2015-0522), Generalitat de Catalunya (Grant No. SGR874 and CERCA program), DFG (FOR2414),  and EU (PCIG13-GA-2013 No. 631633 and H2020-FETPROACT-2014 No. 641122).

\bibliography{MagneticPolaron}

\widetext
\pagebreak
\begin{center}
\textbf{\large Supplementary Materials}
\end{center}

\renewcommand{\theequation}{S\arabic{equation}}
\renewcommand{\thefigure}{S\arabic{figure}}
\renewcommand{\bibnumfmt}[1]{[S#1]}
\setcounter{equation}{0}
\setcounter{figure}{0}

\subsection{Derivation of the effective Hamiltonian}\label{secH}
We first derive the effective Hamiltonian given by Eq.~(4) in the main text. To take into account the initial macroscopic population of the host bosons in the ${\bf k}=0$ mode, we expand $\hat{a}_{{\bf 0}\sigma}$ around $\langle\hat{a}_{{\bf 0}\sigma}\rangle=\sqrt{N_{\rm B}/2}$. Here the factor of $1/2$ accounts for the fact that the bosons are prepared in a superposition of $\uparrow$- and $\downarrow$-states. We then diagonalize the bath Hamiltonian (Eq. (2) in the main text) using the  Bogoliubov transformation:
\eqn{
\hat{a}_{{\bf k},\uparrow}=\frac{1}{\sqrt{2}}\left(\hat{\gamma}_{{\bf k}}^{s}+u_{{\bf k}}\hat{\gamma}_{{\bf k}}^{c}-v_{{\bf k}}\hat{\gamma}_{-{\bf k}}^{\dagger c}\right),\;\;\hat{a}_{{\bf k},\downarrow}=\frac{1}{\sqrt{2}}\left(-\hat{\gamma}_{{\bf k}}^{s}+u_{{\bf k}}\hat{\gamma}_{{\bf k}}^{c}-v_{{\bf k}}\hat{\gamma}_{-{\bf k}}^{\dagger c}\right).
}
Here we introduce the coefficients $u_{\bf k}=\sqrt{(\epsilon_{\bf k}+g_{\rm BB}n_{\rm B})/(2\epsilon_{\bf k}^{c})+1/2}$ and $v_{\bf k}=\sqrt{(\epsilon_{\bf k}+g_{\rm BB}n_{\rm B})/(2\epsilon_{\bf k}^{c})-1/2}$. The resulting expression for the total Hamiltonian $\hat{H}$ of the system is
\eqn{
\hat{H}&=&g_{\rm IB}^{+}n_{\rm B}+\frac{\hat{{\bf P}}^2}{2m_{\rm I}}+\sum_{{\bf k}}\left(\epsilon_{{\bf k}}^{c}\hat{\gamma}_{{\bf k}}^{\dagger c}\hat{\gamma}_{{\bf k}}^{c}+\epsilon_{{\bf k}}^{s}\hat{\gamma}_{{\bf k}}^{\dagger s}\hat{\gamma}_{{\bf k}}^{s}\right)
+\sqrt{\frac{n_{\rm B}}{V}}\!\sum_{{\bf k}}\!\left[g_{\rm IB}^{+}W_{{\bf k}}\!\left(\!\hat{\gamma}_{{\bf k}}^{c}\!+\!\hat{\gamma}_{-{\bf k}}^{\dagger c}\!\right)\!+\!g_{\rm IB}^{-}\!\left(\hat{\gamma}_{{\bf k}}^{s}\!+\!\hat{\gamma}_{-{\bf k}}^{\dagger s}\!\right)\right]e^{-i{\bf k}\hat{{\bf R}}}\nonumber\\
&+&\frac{g_{\rm IB}^{+}}{2V}\sum_{{\bf k},{\bf k}'}\left(V_{{\bf k}{\bf k}'}^{(1)}\hat{\gamma}_{{\bf k}}^{\dagger c}\hat{\gamma}_{{\bf k}'}^{c}e^{i({\bf k}-{\bf k}')\hat{{\bf R}}}+\hat{\gamma}_{{\bf k}}^{\dagger s}\hat{\gamma}_{{\bf k}'}^{s}e^{i({\bf k}-{\bf k}')\hat{{\bf R}}}+V_{{\bf k}{\bf k}'}^{(2)}\hat{\gamma}_{{\bf k}}^{\dagger c}\hat{\gamma}_{{\bf k}'}^{\dagger c}e^{i({\bf k}+{\bf k}')\hat{{\bf R}}}+{\rm H.c.}\right)\nonumber\\
&+&\frac{g_{\rm IB}^{-}}{V}\sum_{{\bf k},{\bf k}'}\left(u_{{\bf k}}\hat{\gamma}_{{\bf k}'}^{\dagger s}\hat{\gamma}_{{\bf k}}^{c}e^{i({\bf k}'-{\bf k})\hat{{\bf R}}}-v_{{\bf k}}\hat{\gamma}_{{\bf k}'}^{\dagger s}\hat{\gamma}_{{\bf k}}^{\dagger c}e^{i({\bf k}'+{\bf k})\hat{{\bf R}}}+{\rm H.c.}\right).
}
To simplify the problem, we transform to the frame comoving with the polaron by using the unitary operator $\hat{U}=e^{i\hat{{\bf R}}\hat{{\bf P}}_{\rm B}}$ with $\hat{{\bf P}}_{B}=\sum_{{\bf k}}{\bf k}(\hat{\gamma}_{{\bf k}}^{\dagger c}\hat{\gamma}_{{\bf k}}^{c}+\hat{\gamma}_{{\bf k}}^{\dagger s}\hat{\gamma}_{{\bf k}}^{s})$ (we set $\hbar=1$). This leads to the effective Hamiltonian $\hat{\cal H}=\hat{U}^{\dagger}\hat{H}\hat{U}$ given by Eq.~(4) in the main text. We note that, in this frame, $\hat{{\bf P}}$ becomes the total momentum of the system and commutes with the effective Hamiltonian $\hat{\cal H}$ and is thus a conserved quantity. We also note that the vertex functions that couple to the spin sector acquire an additional momentum dependence when the SU(2) symmetry of the bath is broken. 
 
 \subsection{The equations of motion and the stationary solution}
 In this section we provide the full expressions for the equations of motion for the variational parameters, given by Eq. (7) in the main text, and derive their stationary solution. As outlined in the main text, the evolution equations for the amplitudes $\alpha_{\bf k}^{c,s}$ are given by the variational condition $\delta[\langle\Psi|i\partial_{t}-\hat{\cal H}|\Psi\rangle]=0$ with respect to the product of coherent states (see Eq.~(6) in the main text). The resulting equations are
\eqn{
i\dot{\alpha}_{{\bf k}}^{c} & = & \Omega_{{\bf k}}^{c}\alpha_{{\bf k}}^{c}-\frac{{\bf k}\cdot({\bf P}-{\bf P}_{\rm B}[\alpha_{{\bf k}}^{c,s}])}{m_{\rm I}}\alpha_{{\bf k}}^{c}+g_{\rm IB}^{+}\left(\sum_{{\bf k}'}V_{{\bf k}{\bf k}'}^{(1)}\alpha_{{\bf k}'}^{c}+\sum_{{\bf k}'}V_{{\bf k}{\bf k}'}^{(2)}\alpha_{{\bf k}'}^{*c}\right)\nonumber \\
 &  & +g_{\rm IB}^{-}\left(u_{{\bf k}}\sum_{{\bf k}'}\alpha_{{\bf k}'}^{s}-v_{{\bf k}}\sum_{{\bf k}'}\alpha_{{\bf k}'}^{*s}\right)+g_{\rm IB}^{+}\sqrt{n_{\rm B}}W_{{\bf k}},\label{eqmotionc}\\
i\dot{\alpha}_{{\bf k}}^{s} & = & \Omega_{{\bf k}}^{s}\alpha_{{\bf k}}^{s}-\frac{{\bf k}\cdot({\bf P}-{\bf P}_{\rm B}[\alpha_{{\bf k}}^{c,s}])}{m_{\rm I}}\alpha_{{\bf k}}^{s}+g_{\rm IB}^{+}\sum_{{\bf k}'}\alpha_{{\bf k}'}^{s}+g_{\rm IB}^{-}\left[\sum_{{\bf k}'}\left(u_{{\bf k}'}\alpha_{{\bf k}'}^{c}-v_{{\bf k}'}\alpha_{{\bf k}'}^{*c}\right)\right]+g_{\rm IB}^{-}\sqrt{n_{\rm B}},\label{eqmotion}
}
where ${\bf P}$ is the total momentum of the system and ${\bf P}_{\rm B}=\sum_{\bf k}{\bf k}(|\alpha^{c}_{\bf k}|^2+|\alpha^{s}_{\bf k}|^2)$.  From the rotational symmetry of the system (for the considered case of a polaron at zero momentum, $\mathbf{P}=0$),  the form of our variational wave function, and its initial condition, it follows that we can set ${\bf P}_{\rm B}={\bf 0}$ in the course of the time evolution. Thus, as expressed in Eq. (7) in the main text, the above equations reduce to  linear inhomogeneous equations for $\alpha_{\bf k}^{c,s}$, where the last terms on the right-hand-side of Eqs. (\ref{eqmotionc}) and (\ref{eqmotion}) describe  driving forces.
 
Next, the stationary solution  $\overline{\alpha}_{\bf k}^{c,s}$ is derived by setting the left-hand-sides of Eqs. (\ref{eqmotionc}) and (\ref{eqmotion}) to zero. From the fact that the real and imaginary parts of the right-hand sides of Eqs. (\ref{eqmotionc}) and (\ref{eqmotion}) must vanish independently, it follows that the imaginary parts of $\overline{\alpha}_{\bf k}^{c,s}$ are zero. Then, by solving the coupled integral equations for the remaining real parts of $\overline{\alpha}_{\bf k}^{c,s}$, we obtain the stationary solution:
\eqn{\label{stsol}
\overline{\alpha}_{{\bf k}}^{c}  =  -\frac{\sqrt{n_{\rm B}}W_{{\bf k}}}{\Omega_{{\bf k}}^{c}}\frac{g_{\rm IB}^{+}(1+g_{\rm IB}^{+}B)-(g_{\rm IB}^{-})^{2}B}{D},\;\;\overline{\alpha}_{{\bf k}}^{s} =  -\frac{\sqrt{n_{\rm B}}}{\Omega_{{\bf k}}^{s}}\frac{g_{\rm IB}^{-}}{D}.
}
Here we introduce
\eqn{
D = (1+g_{\rm IB}^{+}A)(1+g_{\rm IB}^{+}B)-(g_{\rm IB}^{-})^{2}AB,\;\;
A  =  \sum_{{\bf k}}\frac{W_{{\bf k}}^{2}}{\Omega_{{\bf k}}^{c}},\;\;B=\sum_{{\bf k}}\frac{1}{\Omega_{{\bf k}}^{s}}.
}
When the interaction between the impurity and the two-component host bosons are not equal, i.e., $g^{-}_{\rm IB}\neq 0$, the stationary solution  $\overline{\alpha}_{\bf k}^{s}$ is non-zero and the magnetic polaron is formed. 

The energy of the magnetic polaron is given by the expectation value  $E_{\rm mpol}=\langle\Psi_{\rm mpol}|\hat{\cal H}|\Psi_{\rm mpol}\rangle$ with respect to the stationary state  $|\Psi_{\rm mpol}\rangle=\exp\left[{\sum_{{\bf k}}\left(\overline{\alpha}_{{\bf k}}^{c}(t)\hat{\gamma}_{{\bf k}}^{c}+\overline{\alpha}_{{\bf k}}^{s}(t)\hat{\gamma}_{{\bf k}}^{s}-{\rm h.c.}\right)}\right]|0\rangle$. Using the solution (\ref{stsol}) and expressing the interaction strengths $g_{{\rm IB},\sigma}$ in terms of the scattering lengths $a_{{\rm IB},\sigma}$ by the Lippmann-Schwinger equation, 
\eqn{\label{LSeq}
\frac{1}{g_{{\rm IB},\sigma}}=\frac{m_{\rm red}}{2\pi a_{{\rm IB},\sigma}}-\frac{1}{V}\sum_{\bf k}^{\Lambda}\frac{2m_{\rm red}}{{\bf k}^2},
} 
we obtain
\eqn{
E_{\rm mpol}=\frac{2\pi n_{\rm B}}{m_{\rm red}\left(1/a^{+}_{\rm IB}-1/l_{0}\right)},
}
where
\eqn{
a_{\rm IB}^{+}=\frac{a_{{\rm IB},\uparrow}+ a_{{\rm  IB},\downarrow}}{2},\;\; l_{0}=\left(4\pi\sum_{{\bf k}}\frac{1}{{\bf k}^{2}}-\frac{2\pi}{m_{{\rm red}}}\sum_{{\bf k}}\frac{W_{{\bf k}}^{2}}{\Omega_{{\bf k}}^{c}}\right)^{-1}.
}
These expressions are fully regularized and the momentum cutoff $\Lambda$, introduced in Eq.~\eqref{LSeq}, can be taken to infinity.

\subsection{Derivation of the eigenvalue equation}
The energy of the magnetic-dressed bound state  is given by the  eigenvalue equation (9) of the main text. To derive this equation we consider the ansatz  
\eqn{
|\Psi_{{\rm b}}(t)\rangle=\sum_{{\bf k}}\left(\psi_{{\bf k}}^{c}(t)\hat{\gamma}_{{\bf k}}^{\dagger c}+\psi_{{\bf k}}^{s}(t)\hat{\gamma}_{{\bf k}}^{\dagger s}\right)|\Psi_{{\rm mpol}}\rangle.
}
In this state a single phonon and magnon excitation is added to the magnetic polaron $|\Psi_{\rm mpol}\rangle$ which allows to fully account for the underlying two-body bound states. The equations of motion for $\psi_{\bf k}^{c,s}$ are derived from the variational condition $\delta[\langle\Psi_{\rm b}|i\partial_{t}-\hat{\cal H}|\Psi_{\rm b}\rangle]=0$. They are given by
\eqn{\label{eqmotionpc}
i\dot{\psi}_{{\bf k}}^{c} & = & (E_{{\rm pol}}+\Omega_{{\bf k}}^{c})\psi_{{\bf k}}^{c}+g_{\rm IB}^{+}\sum_{{\bf k}'}V_{{\bf k}{\bf k}'}^{(1)}\psi_{{\bf k}'}^{c}+g_{\rm IB}^{-}u_{{\bf k}}\sum_{{\bf k}'}\psi_{{\bf k}'}^{s},\\
i\dot{\psi}_{{\bf k}}^{s} & = & (E_{{\rm pol}}+\Omega_{{\bf k}}^{s})\psi_{{\bf k}}^{s}+g_{\rm IB}^{+}\sum_{{\bf k}'}\psi_{{\bf k}'}^{s}+g_{\rm IB}^{-}\sum_{{\bf k}'}u_{{\bf k}'}\psi_{{\bf k}'}^{c}.\label{eqmotionps}
}
In order to find the  eigenmodes of these equations, we assume the  solutions of the form $\psi_{\bf k}^{c,s}\propto e^{-i\omega t}$ which oscillate in time with frequency $\omega$. Substituting this ansatz into Eqs. (\ref{eqmotionpc}) and (\ref{eqmotionps}), we obtain the  equation
\eqn{\label{matrix}
\begin{pmatrix}
1-\frac{ g_{{\rm IB}}^{+}}{2}\Pi_{w^2}&-\frac{ g_{{\rm IB}}^{+}}{2}\Pi&0&- g_{{\rm IB}}^{-}\Pi_{uw}\\
-\frac{ g_{{\rm IB}}^{+}}{2}\Pi &1-\frac{ g_{{\rm IB}}^{+}}{2}\Pi_{w^{-2}}&0&- g_{{\rm IB}}^{-}\Pi_{uw^{-1}}\\
-\frac{ g_{{\rm IB}}^{+}}{2}\Pi_{uw}&-\frac{ g_{{\rm IB}}^{+}}{2}\Pi_{uw^{-1}}&1&- g_{{\rm IB}}^{-}\Pi_{u^2}\\
0&0&- g_{{\rm IB}}^{-}\Pi_{s}&1- g_{{\rm IB}}^{+}\Pi_{s}\\
\end{pmatrix}
\begin{pmatrix}
\sum_{\bf k}W_{\bf k}\psi_{\bf k}^{c}\\
\sum_{\bf k}W_{\bf k}^{-1}\psi_{\bf k}^{c}\\
\sum_{\bf k}u_{\bf k}\psi_{\bf k}^{c}\\
\sum_{\bf k}\psi_{\bf k}^{s}\\
\end{pmatrix}
=
\begin{pmatrix}
0\\
0\\
0\\
0\\
\end{pmatrix},
}
where we define
\eqn{
\Pi_{w^{\pm 2}}&=&\sum_{\bf k}\frac{W_{\bf k}^{\pm 2}}{\omega-E_{\rm mpol}-\Omega_{\bf k}^{c}},\;\; \Pi=\sum_{\bf k}\frac{1}{\omega-E_{\rm mpol}-\Omega_{\bf k}^{c}},\;\;\Pi_{uw^{\pm 1}}=\sum_{\bf k}\frac{u_{\bf k}W_{\bf k}^{\pm 1}}{\omega-E_{\rm mpol}-\Omega_{\bf k}^{c}}, \nonumber\\
&&\Pi_{u^2}=\sum_{\bf k}\frac{u_{\bf k}^2}{\omega-E_{\rm mpol}-\Omega_{\bf k}^{c}},\;\;\Pi_{s}=\sum_{\bf k}\frac{1}{\omega-E_{\rm mpol}-\Omega_{\bf k}^{s}}.
}
Equation (\ref{matrix}) has nontrivial solutions only if the determinant of the matrix on the left-hand-side vanishes. Expressing the interaction strengths $g_{{\rm IB},\sigma}$ in terms of the scattering lengths $a_{{\rm IB},\sigma}$ via Eq. (\ref{LSeq}) and collecting the leading terms in the limit of $\Lambda\to\infty$, we obtain the  equation (see Eq.~(9) in the main text)

\eqn{
\left[a_{\rm IB}^{+}-\left(4\pi\sum_{\bf k}^{\Lambda}\frac{1}{{\bf k}^2}+\frac{2\pi}{m_{\rm red}}\Pi_{s}\right)^{-1}\right]\left[a_{\rm IB}^{+}-\left(4\pi\sum_{\bf k}^{\Lambda}\frac{1}{{\bf k}^2}+\frac{2\pi}{m_{\rm red}}\frac{\Pi_{w^2}+\Pi_{w^{-2}}}{2}\right)^{-1}\right]=\left(a_{\rm IB}^{-}\right)^2.
}
Solving this equation for $\omega$  gives the bound state energy $\omega_{\rm bound}$.

\subsection{Asymptotic scaling of a generation of spin excitations}
As shown in the panel II in Fig. 2 in the main text, the number of spin-waves generated obeys an asymptotic $\sqrt{t}$ behavior at long times. This behavior can be understood from a simple scaling argument. For the sake of simplicity, let us neglect interactions between different momentum modes and consider the simplified Hamiltonian where different sectors of momentum $\bf k$ are decoupled: $\hat{H}_{\bf k}=\omega_{\bf k}\hat{\gamma}_{\bf k}^{\dagger}\hat{\gamma}_{\bf k}+V_{\bf k} (\hat{\gamma}_{\bf k}+\hat{\gamma}_{\bf k}^{\dagger})$. Here $\hat{\gamma}_{\bf k}$ denotes a spin or charge annihilation operator at momentum $\bf k$ and we assume $V_{-{\bf k}}=V_{{\bf k}}$. In the non-equilibrium problem such as the one studied in this work, the number of excitations $n_{\bf k}$ in the mode $\bf k$ will in general oscillate. However, for times larger than $t>1/\omega_{k}$, one can give a simple scaling argument for the behavior of $n_{\mathbf k}$. At such times, the occupation of excitations have a scaling $\langle\hat{\gamma}_{\bf k}\rangle\propto V_{\bf k}/\omega_{\bf k}$ and thus $n_{\bf k}\propto (V_{\bf k}/\omega_{\bf k})^2$ as can be seen from an inspection of the equation of motion.  

Let us then consider the total number of excitations at time $t$. We identify all modes ${\bf k}$ satisfying $\omega_{{\bf k}}t>1$, i.e., the modes $|{\bf k}|>k^{*}$ where $k^{*}$ is determined by $\omega_{k^*}t=1$, as contributions to the excitations. Using the estimate $n_{\bf k}\propto (V_{\bf k}/\omega_{\bf k})^2$, we then integrate over these modes to find  for the spin excitations
\eqn{\label{scalingx}
N^{s}(t)=\int d^{3}{\bf k}\;n^{s}_{\bf k}\propto \int_{k^{*}}^{\infty} k^2 dk\;\frac{1}{k^4} \propto \sqrt{t}.
} 
Here we use the fact that  for the spin sector $V_{\bf k} = {\rm const.}$, and the magnon dispersion relation scales as $\omega_{\bf k}\propto k^2$, leading to $k^{*}\propto 1/\sqrt{t}$. We emphasize that collective excitations satisfying these simple scalings can ubiquitously appear in a variety of systems such as  multi-component Bose-Einstein condensates \cite{AVG09,MGE14}, fermionic gases \cite{MA16,VG17}, and Rydberg or dipolar gases \cite{ZJ16,FB16,DP17}. 

In contrast, for charge excitations, one has $V_{\bf k}\propto \sqrt{k}$ and the dispersion is linear $\omega_{\bf k}\propto k$ at small $\bf k$, leading to $k^{*}\propto 1/t$ and $n_{\bf k}\propto 1/k$. Hence, if integrated from $k^*$ to infinity, the total number of charge excitations diverges in the ultraviolet (UV) limit. This suggests that the main contribution to $N^{c}$ comes from the UV limit. At such a large momentum, however, we can use the same scaling argument as for the spin excitations. Since this gives a convergent result in the UV limit, we expect that the number of charge excitations should soon reach a constant number and saturate, in  agreement with our numerical findings.  

We note that the $\sqrt{t}$ behavior remains observable also  in the presence of a small imbalance in the boson-boson scattering length (i.e.,~broken SU(2) symmetry). In this case, the magnon dispersion relation has a linear low-energy contribution and the number of spin excitations will ultimately saturate.  However, since the imbalance in scattering length is typically very small, we expect a large time window for which the $\sqrt{t}$ behavior remains valid before saturation. Specifically, let $t=t^*$ be the time scale at which $k^*$ (determined by $\omega_{k^*} t =1$) reaches a small value such that the magnon dispersion becomes linear for $k<k^*$. Under this condition, the scaling argument given in Eq.~(\ref{scalingx}) remains valid as long as $t<t^*$. Hence we expect that  $t^*$ is long enough to experimentally  observe the $\sqrt{t}$ behavior.

\subsection{Coupled oscillators dynamics in the presence of two bound states}

\begin{figure}[t]
\includegraphics[width=120mm]{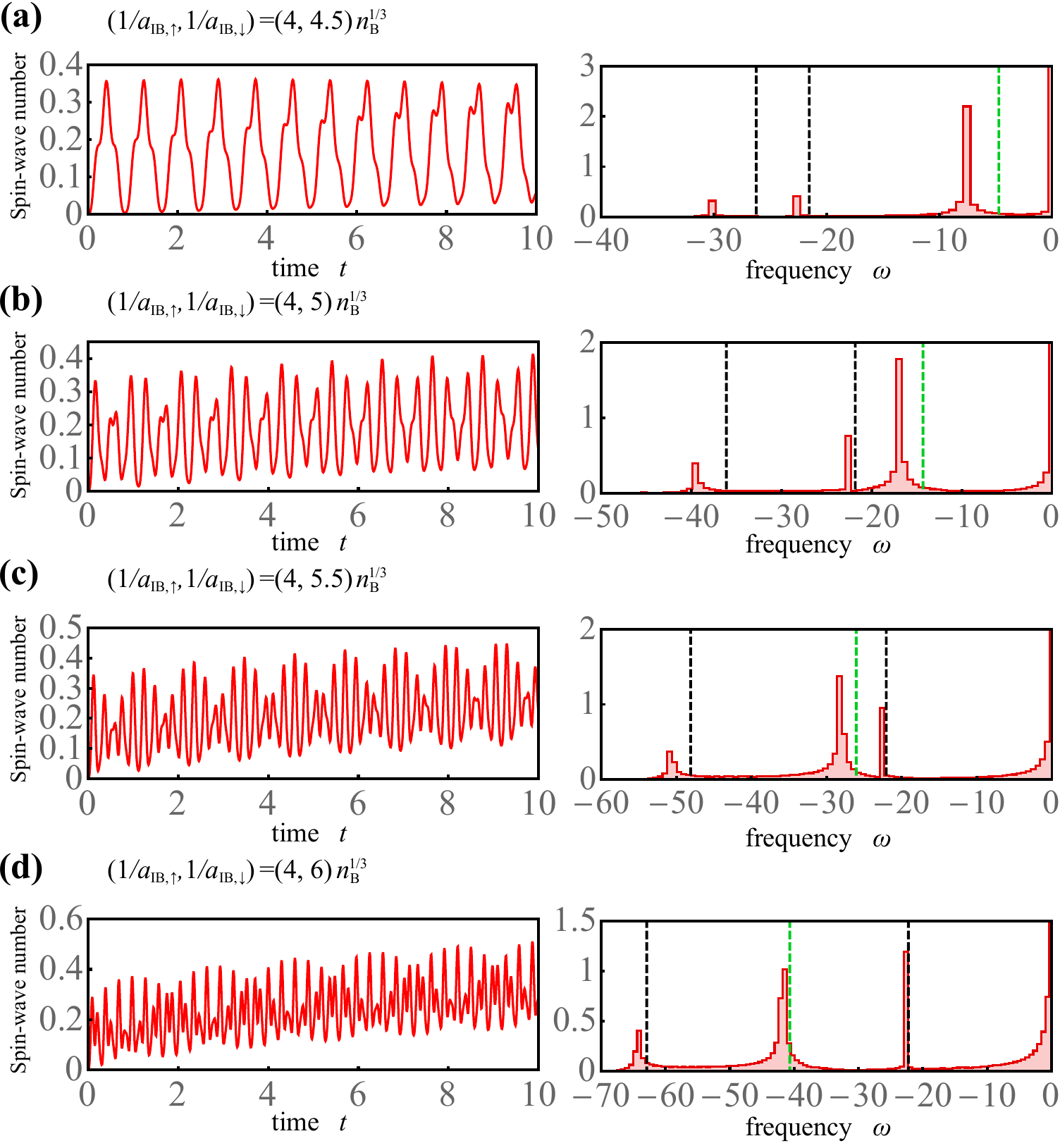}
 \caption{\label{figs2}
Dynamics of bath spins and their Fourier spectra in the presence of two bound states. One of the scattering lengths is set to $1/(a_{{\rm IB},\uparrow}n_{\rm B}^{1/3})=4$ while  the other ($1/(a_{{\rm IB},\downarrow}n_{\rm B}^{1/3})$) is varied in the range from 4.5 to 6 [(a)-(d)].  The black, dashed lines in the Fourier spectra indicate the eigenenergies $\omega_{\rm bound}$ of the magnetic bound states calculated from Eq. (9) of the main text, while the green dashed line show the difference of these energies. We use the parameters $a_{\rm BB}n_{\rm B}^{1/3}=0.05$ and $m_{\rm I}/m_{\rm B}=0.95$. Time and frequency are shown in  units of $m_{\rm B}/(\hbar n_{\rm B}^{2/3})$. 
}
\end{figure}

When two bound states are present (defining the regime III shown in Fig. 1(b) of the main text), the spin and charge excitations exhibit multi-frequency oscillations. Remarkably, continuum modes in the bath mediate a coupling between the two bound states. As a consequence,  the oscillation frequencies found in the spin dynamics are shifted from the ``bare" eigenfrequencies that are determined  by the magnetic bound state energies given by Eq. (9) in the main text.

To clarify this point further, we show  in Fig. \ref{figs2}  the dynamics of bath spins for varying scattering lengths. When both  states are weakly bound and close in energy (i.e., both scattering lengths are large and take on similar values), the coupling between the two bound states can be strong enough to induce shifts in the oscillation frequencies as shown in Fig. \ref{figs2}(a). In this regime, a large peak placed at the difference of the two oscillatory modes indicates a strong coupling of the two  bound states. As  the strength of one of the scattering lengths is decreased, while keeping the other unchanged, the coupling of the two bound states becomes weak and the oscillation frequencies eventually converge to the `bare' bound state energies indicated by the dashed black lines (see Fig.~\ref{figs2}(b)-(d)). As the energy gap between both states increases,  the gradual decoupling of the two bound states can also be seen as a decrease of the peak height at the difference between the two oscillatory mode energies.

\subsection{Decoherence induced by a difference in the scattering lengths of host bosons}
In the main text, we assumed that the scattering length $a_{\rm BB}$ between host bosons is independent of the spin components $\sigma=\uparrow,\downarrow$. However, in practice, there exists typically a small imbalance in the boson-boson interactions. For example, identifying the hyperfine states $|F=1,m_{F}=1\rangle$ and $|F=1,m_{F}=0\rangle$ as $\uparrow$-and $\downarrow$-state, respectively, the imbalance in scattering lenghts is $\sim 0.4\%$ for $^{41}$K \cite{TomzaPriv} and $\sim 0.5\%$ for $^{87}$Rb atoms \cite{Kempen02}. In general, this weak symmetry breaking causes  spin decoherence that is additional to  the one induced by the impurity. To estimate the size of such a contribution, we consider the Hamiltonian of a two-component gas of host bosons in absence of the impurity 
\eqn{
\hat{H}_{\rm B}&=&\sum_{{\bf k}\sigma}\!\epsilon_{\bf k}\hat{a}_{{\bf k}\sigma}^{\dagger}\!\hat{a}_{{\bf k}\sigma}\nonumber\\
&+&\frac{1}{2V}\!\!\sum_{{\bf k},{\bf k}',{\bf q}}\!\!\left(g_{{\rm BB}\uparrow\uparrow}\hat{a}_{{\bf k}\!+\!{\bf q},\uparrow}^{\dagger}\hat{a}_{{\bf k}'-{\bf q},\uparrow}^{\dagger}\hat{a}_{{\bf k},\uparrow}\hat{a}_{{\bf k}',\uparrow}\!+\!2g_{{\rm BB}\uparrow\downarrow}\hat{a}_{{\bf k}+{\bf q},\uparrow}^{\dagger}\hat{a}_{{\bf k}'-{\bf q},\downarrow}^{\dagger}\hat{a}_{{\bf k},\uparrow}\hat{a}_{{\bf k}',\downarrow}\!+\!g_{{\rm BB}\downarrow\downarrow}\hat{a}_{{\bf k}\!+\!{\bf q},\downarrow}^{\dagger}\hat{a}_{{\bf k}'-{\bf q},\downarrow}^{\dagger}\hat{a}_{{\bf k},\downarrow}\hat{a}_{{\bf k}',\downarrow}\right).
}
Here $g_{{\rm BB}\sigma\sigma'}$ denotes the interaction strength between host bosons of spin component $\sigma$ and $\sigma'$. Let us introduce imbalance parameters $r_{1,2}$ by
\eqn{
g_{{\rm BB}\uparrow\uparrow}=g_{{\rm BB}\uparrow\downarrow}(1+r_{1}),\; g_{{\rm BB}\downarrow\downarrow}=g_{{\rm BB}\uparrow\downarrow}(1+r_{2}),\; g_{{\rm BB}\uparrow\downarrow}\equiv g_{{\rm BB}}.
}
Then, by following a similar procedure as outlined in Sec.~\ref{secH}, we can diagonalize the Hamiltonian to obtain
\eqn{\label{Himbalance}
\hat{H}_{\rm B}=\sum_{{\bf k}}\left(\tilde{\epsilon}_{{\bf k}}^{s}\hat{\tilde{\gamma}}_{{\bf k}}^{\dagger s}\hat{\tilde{\gamma}}_{{\bf k}}^{s}+\tilde{\epsilon}_{{\bf k}}^{c}\hat{\tilde{\gamma}}_{{\bf k}}^{\dagger c}\hat{\tilde{\gamma}}_{{\bf k}}^{c}\right),
}
where the dispersion relations are given by
\eqn{
\tilde{\epsilon}_{{\bf k}}^{s,c}=\sqrt{\left(\frac{{\bf k}^{2}}{2m_{\rm B}}\right)^{2}+\frac{gn_{\rm B}{\bf k}^{2}}{2m_{\rm B}}\left[\left(1+\frac{r_{1}+r_{2}}{2}\right)\mp\sqrt{1+\left(\frac{r_{1}-r_{2}}{2}\right)^2}\right]}.
}
To ensure that the energies are real, we require that the parameters satisfy the miscible condition:
\eqn{
(1+r_{1})(1+r_{2})>1 \iff g_{{\rm  BB}\uparrow\uparrow}g_{{\rm BB}\downarrow\downarrow}>g_{{\rm BB}\uparrow\downarrow}^{2}.
}
The operators $\hat{\tilde{\gamma}}_{{\bf k}}^{s,c}$ are related to $\hat{a}_{{\bf k},\sigma}$ by a Bogoliubov transformation. As an example, we show the expressions in the case of $r_{1}=r_{2}=r$:
\eqn{\label{Bogoliubov}
\begin{cases}
\hat{a}_{{\bf k}\uparrow}=\frac{1}{\sqrt{2}}\left(\tilde{u}_{{\bf k}}^{s}\hat{\tilde{\gamma}}_{{\bf k}}^{s}-\tilde{v}_{-{\bf k}}^{s}\hat{\tilde{\gamma}}_{-{\bf k}}^{\dagger s}+\tilde{u}_{{\bf k}}^{c}\hat{\tilde{\gamma}}_{{\bf k}}^{c}-\tilde{v}_{-{\bf k}}^{c}\hat{\tilde{\gamma}}_{-{\bf k}}^{\dagger c}\right)\\
\hat{a}_{{\bf k}\downarrow}=\frac{1}{\sqrt{2}}\left(-\tilde{u}_{{\bf k}}^{s}\hat{\tilde{\gamma}}_{{\bf k}}^{s}+\tilde{v}_{-{\bf k}}^{s}\hat{\tilde{\gamma}}_{-{\bf k}}^{\dagger s}+\tilde{u}_{{\bf k}}^{c}\hat{\tilde{\gamma}}_{{\bf k}}^{c}-\tilde{v}_{-{\bf k}}^{c}\hat{\tilde{\gamma}}_{-{\bf k}}^{\dagger c}\right),
\end{cases}
}
where 
\eqn{
\tilde{u}_{{\bf k}}^{s}&=&\sqrt{\frac{\frac{{\bf k}^{2}}{2m_{\rm B}}+\frac{gn_{\rm B}}{2}r}{2\sqrt{\left(\frac{{\bf k}^{2}}{2m_{\rm B}}\right)^{2}+\frac{gn_{\rm B}{\bf k}^{2}}{2m_{\rm B}}r}}+\frac{1}{2}},\nonumber\\
\tilde{v}_{{\bf k}}^{s}&=&\frac{gn_{\rm B}r}{2\sqrt{2}}\left[\left(\frac{{\bf k}^{2}}{2m_{\rm B}}\right)^{2}+\frac{gn_{\rm B}{\bf k}^{2}}{2m_{\rm B}}r+\left(\frac{{\bf k}^{2}}{2m_{\rm B}}+\frac{gn_{\rm B}}{2}r\right)\sqrt{\left(\frac{{\bf k}^{2}}{2m_{\rm B}}\right)^{2}+\frac{gn_{\rm B}{\bf k}^{2}}{2m_{\rm B}}r}\right]^{-1/2}.
}
For simplicity, we focus on this case in the following. A generalization to the case of  $r_{1}\neq r_{2}$ is straightforward and leads to the same qualitative physics.

In order to study  how the initially prepared superposition state $|\Psi_{\rm BEC}\rangle\propto(\hat{a}^{\dagger}_{{\bf 0}\uparrow}+\hat{a}^{\dagger}_{{\bf 0}\downarrow})^{N_{\rm B}}|0\rangle$ dephases due to the imbalance in the spin-dependent boson-boson scattering lengths, we consider  the initial state in terms of the operators $\hat{\tilde{\gamma}}_{{\bf k}}^{s,c}$:
\eqn{
|\Psi_{\rm BEC}(0)\rangle\propto\exp\left[\frac{1}{2}\sum_{{\bf k}\neq0}\left(\frac{\tilde{v}_{-{\bf k}}^{s}}{\tilde{u}_{{\bf k}}^{s}}\hat{\tilde{\gamma}}_{{\bf k}}^{\dagger s}\hat{\tilde{\gamma}}_{-{\bf k}}^{\dagger s}+\frac{\tilde{v}_{-{\bf k}}^{c}}{\tilde{u}_{{\bf k}}^{c}}\hat{\tilde{\gamma}}_{{\bf k}}^{\dagger c}\hat{\tilde{\gamma}}_{-{\bf k}}^{\dagger c}\right)\right]|0\rangle_{\gamma},
}
which satisfies $\hat{a}_{{\bf k}\sigma}|\Psi_{\rm BEC}(0)\rangle=0$ for ${\bf k}\neq 0$, where $|0\rangle_{\gamma}$ denotes the vacuum of the $\hat{\tilde{\gamma}}_{\bf k}^{s,c}$ operators. From the Hamiltonian (\ref{Himbalance}), the time evolution of the quantum state follows 
\begin{figure}[t]
\includegraphics[width=145mm]{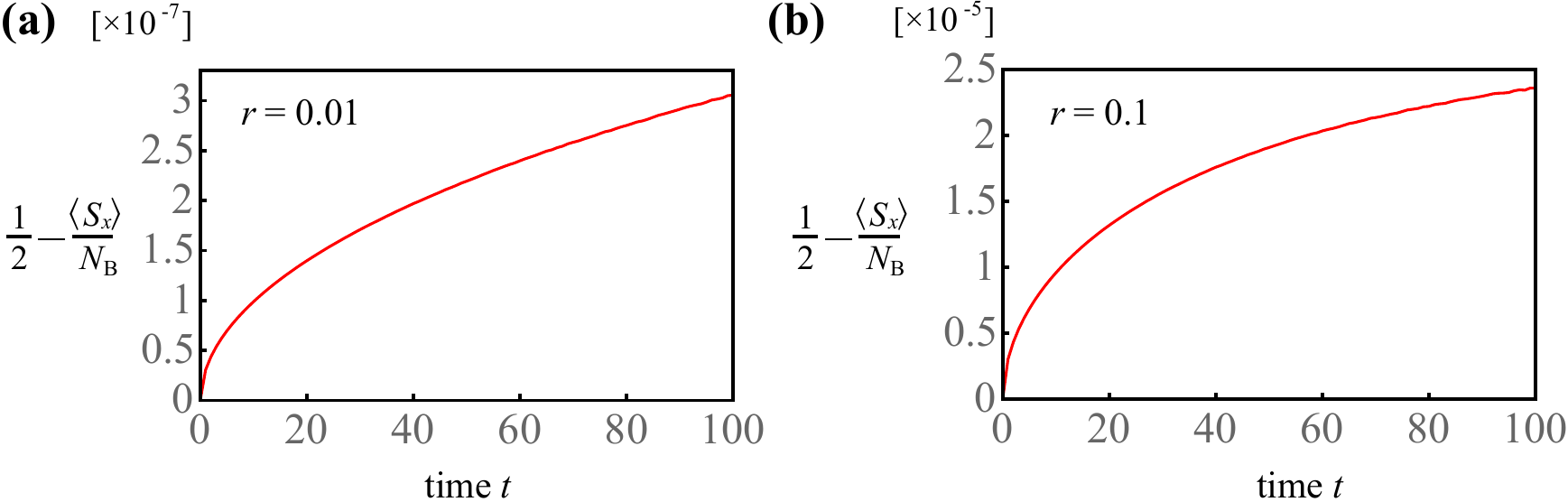}
 \caption{\label{figs1}
The decoherence induced by an imbalance in the interactions between the different components of the host bosons. The time evolutions are plotted for (a) $r=0.01$ and (b) $r=0.1$. The boson-boson interaction is set to  $a_{\rm BB}n^{1/3}_{\rm B}=0.1$ and we plot time  in units of $m_{\rm B}/(\hbar n_{\rm B}^{2/3})$.
}
\end{figure}
 \eqn{
 |\Psi_{\rm BEC}(t)\rangle\propto\exp\left[\frac{1}{2}\sum_{{\bf k}\neq0}\left(\frac{\tilde{v}_{-{\bf k}}^{s}e^{-2i\tilde{\epsilon}_{{\bf k}}^{s}t}}{\tilde{u}_{{\bf k}}^{s}}\hat{\tilde{\gamma}}_{{\bf k}}^{\dagger s}\hat{\tilde{\gamma}}_{-{\bf k}}^{\dagger s}+\frac{\tilde{v}_{-{\bf k}}^{c}e^{-2i\tilde{\epsilon}_{{\bf k}}^{c}t}}{\tilde{u}_{{\bf k}}^{c}}\hat{\tilde{\gamma}}_{{\bf k}}^{\dagger c}\hat{\tilde{\gamma}}_{-{\bf k}}^{\dagger c}\right)\right]|0\rangle_{\gamma}.
 }
 Then, by denoting $\langle\cdots\rangle$ as the expectation value with respect to $|\Psi_{\rm BEC}(t)\rangle$, the time evolution of the spin operator becomes
 \eqn{
\frac{\langle\hat{S}_{x}\rangle}{ N_{\rm B}}& = & \left\langle \frac{1}{2N_{\rm B}}\sum_{{\bf k}}\left(\hat{a}_{{\bf k},\uparrow}^{\dagger}\hat{a}_{{\bf k},\downarrow}+\hat{a}_{{\bf k},\downarrow}^{\dagger}\hat{a}_{{\bf k},\uparrow}\right)\right\rangle \\
 & = & \frac{1}{2}-\frac{1}{n_{\rm B}}\int\frac{d^{3}{\bf k}}{(2\pi)^{3}}2(\tilde{u}_{{\bf k}}^{s})^{2}(\tilde{v}_{{\bf k}}^{s})^{2}(1-\cos(2\tilde{\epsilon}_{{\bf k}}^{s}t)),\label{decoherence}
 }
 where we used the expressions of the Bogoliubov transformations (\ref{Bogoliubov}).
The second term in Eq.~(\ref{decoherence}) represents the decoherence factor induced by the spin-dependent internal interactions between the host bosons. The integral over $(\tilde{u}_{\bf k}^{s})^2$ roughly equals the number of excited particles, which is typically less than $1\%$, and $\tilde{v}_{\bf k}^{s}$ is on the order of $r$. Thus, the decoherence factor can  be estimated by the multiplication of these two factors.  

As an example,  we assume an imbalance $r=0.01$. Then the total decoherence factor induced by the internal dynamics (\ref{decoherence}) is about $< 10^{-6}$ which is negligible  compared to the dephasing induced by the impurity. Figure \ref{figs1} shows the time evolution of the decoherence given by Eq.~(\ref{decoherence}) for the imbalance parameters $r=0.01$, and $0.1$. Our numerical finding supports the above estimate of the decoherence factor. In particular, the decoherence is still greatly suppressed even for an  imbalance in the boson-boson scattering lengths of about 10\%. Thus, our predictions on the magnetic polaron dynamics studied in the main text should be  detectable also using a miscible pair of hyperfine states of $^{23}$Na by identifying, for instance, $|F=1,m_{F}=1\rangle$ as the $\uparrow$-state and $|F=1,m_{F}=0\rangle$ as the $\downarrow$-state, leading to an imbalance $\sim 8\%$ \cite{SC00}. For this choice, $^{40}$K  \cite{WC12} will be the most promising candidate for the  impurity atoms.

\end{document}